# Phase separation in LuFeO$_3$ films

Shi Cao[1*], Xiaozhe Zhang[1,2], Kishan Sinha[1], Wenbin Wang[3], Jian Wang[4], Peter A. Dowben[1] and Xiaoshan Xu[1*]

[1]Department of Physics and Astronomy, Nebraska Center for Materials and Nanoscience, University of Nebraska, Lincoln, Nebraska 68588, USA

[2]Department of Physics, Xi'an Jiaotong University, Xi'an 710049, China

[3]Department of Physics, Fudan University, Shanghai 200433, China

[4]Canadian Light Source Inc., 44 Innovation Boulevard, Saskatoon, SK S7N 2V3, Canada

**Abstract:**

The structural transition at about 1000 °C, from the hexagonal to the orthorhombic phase of LuFeO$_3$, has been investigated in thin films of LuFeO$_3$. Separation of the two structural phases of LuFeO$_3$ occurs on a length scale of micrometer, as visualized in real space using X-ray photoemission electron microscopy (X-PEEM). The results are consistent with X-ray diffraction and atomic force microscopy obtained from LuFeO$_3$ thin films undergoing the irreversible structural transition from the hexagonal to the orthorhombic phase of LuFeO$_3$, at elevated temperatures. The sharp phase boundaries between the structural phases are observed to align with the crystal planes of the hexagonal LuFeO$_3$ phase. The coexistence of different structural domains indicates that the irreversible structural transition, from the hexagonal to the orthorhombic phase in LuFeO$_3$, is a first order transition, for epitaxial hexagonal LuFeO$_3$ films grown on Al$_2$O$_3$.

* xiaoshan.xu@unl.edu, caoshi86@gmail.com

Ferroelectricity and ferromagnetism are foundations of numerous technologies. The combination of ferroelectricity and ferromagnetism, namely multiferroicity, is believed to have great importance for future technologies, although very few materials are known to be ferroelectric and ferromagnetic at the same time. Hexagonal LuFeO$_3$ (h-LuFeO$_3$) is a multiferroic material that exhibits spontaneous electric and magnetic polarizations simultaneously.[1–5] It has been predicted that the magnetic dipole moment in h-LuFeO$_3$ can be switched by an electric field,[6] which is appealing for application in energy efficient information storage and processing.[7–9]

Rather than h-LuFeO$_3$, the orthorhombic crystallographic structure (o-LuFeO$_3$) is the thermodynamically stable bulk structure of LuFeO$_3$ with standard conditions,[10] meaning that the free energy of o-LuFeO$_3$ is lower than that of the h-LuFeO$_3$. In epitaxial thin films, the film-substrate interfacial energy may favor the h-LuFeO$_3$ structure, if the symmetry of the substrate is triangular or hexagonal. This effect can stabilize the hexagonal structure in epitaxial thin films, to a certain critical thickness,[1,11–14] at which point the lower free energy of o-LuFeO$_3$ dominates. Beyond the critical thickness, h-LuFeO$_3$ films may exist as a metastable state, because of an energy barrier to nucleate the orthorhombic phases within the hexagonal phase. In this case, at elevated temperatures, a transition from hexagonal to orthorhombic structural phases occurs, as the thermal energy will increasingly overcome the energy barrier.

In this work, we have studied the transition from the hexagonal to orthorhombic phase in h-LuFeO$_3$ films, grown on Al$_2$O$_3$ (0001) substrates. We found that in h-LuFeO$_3$ films, the transition occurs at around 1000 °C, with a coexistence of the two structural phases. The structural phase separation was observed on the micrometer scale; the boundaries between the two phases are aligned with the crystal planes of the h-LuFeO$_3$ phase. These findings suggest a minimal stability problem of h-LuFeO$_3$ films for application, and a self-organization of the sharp hexagonal/orthorhombic interface that involves a strong magnetic order (o-LuFeO$_3$, $T_N$ = 620 K)[15] and a strong ferroelectric order (h-LuFeO$_3$, $T_C$=1050 K).[1]

Hexagonal LuFeO$_3$ (001) films were grown on Al$_2$O$_3$ (0001) substrates using pulsed laser deposition at 750 °C in a 5 mtorr oxygen environment.[1,16–18] To test the thermal stability of hexagonal LuFeO$_3$, we carried out a sequence of annealing on a film sample of ~40 nm thickness. For each annealing step, the temperature was raised slowly (5 °C/min) from room temperature to the annealing temperature ($T_A$) and then annealed at that temperature for 3 hours, followed by a slow cooldown (5 °C/min) to room temperature. After each annealing cycle, X-ray diffraction (XRD) was taken on the sample, using a Rigaku D/Max-B diffractometer, with the Co K$\alpha$ radiation (1.7903 Å). The annealing/XRD sequence was repeated for 8 temperatures, in the order: 600, 700, 800, 850, 900, 950, 1000, and 1050 °C. Another heated sample (~10 nm thickness) was studied using atomic force microscopy (AFM) using a TT-AFM from AFM Workshop. This same sample was also investigated by X-ray absorption spectroscopy (XAS) studies through the X-ray photoemission electron microscope (X-PEEM), at the SM beamline of the Canadian Light Source (CLS) with linearly polarized X-rays at room temperature in ultrahigh vacuum; the X-ray beam incident angle was 16 degree[17]. The XAS was obtained by pixel-by-pixel integration of X-PEEM image as a function of photon energy.

By annealing the samples at higher temperatures, we found that the transition starts at about 1000 °C, with clear indications of phase coexistence. Figure 1 displays the XRD pattern of the h-LuFeO$_3$ film right after the growth (as-grown) and after being annealed at different temperatures ($T_A$). Here we use the pseudo cubic unit cell for indexing the o-LuFeO$_3$ diffraction peaks. The film appears to be stable at least up to 700 °C, because the X-ray diffraction patterns are characteristic of the XRD for the as-grown h-LuFeO$_3$ film. This is consistent with our previous result that impurity phase generated at the surface by sputtering may be converted back to the h-LuFeO$_3$ phase, by annealing the sample at 600 °C.[18] Upon increasing $T_A$ above 700 °C, the XRD intensity of the h-LuFeO$_3$ (002) and (004) peaks decreases, and reaches a minimum at $T_A$ = 850 °C, but then increases until diminishing again in the region of 1000 °C, only to disappear at slightly higher temperature of $T_A$ = 1050 °C. The characteristic XRD features of o-LuFeO$_3$ starts to appear at $T_A$ = 1000 °C, at a temperature where the h-LuFeO$_3$ XRD peaks are still present. Because we use the pseudo cubic unit cell for indexing the o-LuFeO$_3$ diffraction peaks, the (111) peak actually corresponds to three different peaks in orthorhombic structure, as seen in Fig. 1.

These results suggest the following scenario for the transition from the h-LuFeO$_3$ phase to the o-LuFeO$_3$ phase. At about 800 °C, conversion to o-LuFeO$_3$ phase occurs locally, but only as structural fluctuations. We posit that in the region of 800 °C, the interfacial energy between the h-LuFeO$_3$ phase and the o-LuFeO$_3$ phase generates a large energy barrier to the nucleation of o-LuFeO$_3$ domains. Thus no indication of o-LuFeO$_3$ phase can be clearly observed in the XRD of the LuFeO$_3$ thin films when quenched back to room temperature, although it is clear that defects and/or dislocations frozen into the h-LuFeO$_3$ thin film degrade the XRD peak intensities dramatically. At higher temperature (1000 °C), the thermal energy is large enough to overcome the energy barrier for the nucleation of the o-LuFeO$_3$ phase; this leads to the separation of the two structural phases into large structural domains. The defects previously frozen into the h-LuFeO$_3$ thin film are now annealed out. As a result, the diffractions signatures of both the hexagonal and orthorhombic phases are now evident. This scenario is also consistent with the dependence of the rocking curve width of the h-LuFeO$_3$ (004) peak on $T_A$. As shown in Fig. 1 inset, the rocking curve width reaches a maximum at $T_A$=850 °C, indicating that the in-plane correlation of the atomic positions is at a minimum, which agrees with peak intensity minimum at $T_A$=850 °C.

The coexistence of the h-LuFeO$_3$ phase and the o-LuFeO$_3$ phase indicates that the transition from the h-LuFeO$_3$ phase to the o-LuFeO$_3$ phase is first order, due to the difference between the densities of the two phases.[19] To verify the existence of the phase separation in real space, and to probe the length scale of the phase separation, we employed atomic force microscopy and X-PEEM on another sample (10 nm) rather than the sample of greater thickness (and thus more suitable for XRD).

Phase separation occurs on a length scale of micrometer, may be visualized, in real space, using X-ray photoemission electron microscopy (X-PEEM). As recently demonstrated,[16,17] the X-ray absorption spectra of the hexagonal and orthorhombic phases are dramatically different, both in spectral shape and in linear dichroism, and these differences in the X-ray absorption spectra (XAS) may be used to distinguish the two phases.[16,17] X-PEEM technique, employing an X-ray source and a high-resolution electron microscope, measures the X-ray absorption spectra with spatial resolution,[20] so it is an ideal technique to conclusively distinguish the two structural phases in real space, by comparing the X-ray absorption spectra at various spatial locations point by point.

Figure. 2(a) presents the X-ray absorption spectra at Fe $L_{III}$ and $L_{II}$ edges for h-LuFeO$_3$ and o-LuFeO$_3$ using linearly polarized X-rays. In hexagonal structure, Fe 3d states split into three irreducible representations $e''$ (xz, yz), $e'$ (xx-yy, xy) and $a_1'$ (zz).[16,18] The energies of these crystal field states follow the order $E_{a_1'} > E_{e'} > E_{e''}$.[17] As shown in Fig. 2(a), for h-LuFeO$_3$, the XAS $e'$ peak (at about 709.5 eV) will only be present with *s* polarization (i.e. with in-plane linearly polarized light),[18] due to the applicable spectroscopic selection rules.[21] For o-LuFeO$_3$, the absorption spectra always show two peaks, in the region of 708 to 712 eV, corresponding to the $e_g$ and $t_{2g}$ crystal field states, independent of polarization of X-ray.[16,17] Therefore, there is a clear correlation between lattice structure and X-ray absorption spectra in LuFeO$_3$.[18] In particular, with *s*-polarized X-ray, the difference between the absorption spectra of h-LuFeO$_3$ and o-LuFeO$_3$ is significant, an aid for distinguishing the two structural phases.

We have used the large contrast obtained in X-PEEM to distinguishing the structural phases using their corresponding difference in electronic structures. Figure. 2(b) shows an X-PEEM image, in a 50 micrometer field of view, taken at photon energy of 709 eV using *s*-polarized X-rays. The contrast in the image can be identified as having an origin in the XAS spectroscopic differences of h-LuFeO$_3$ and o-LuFeO$_3$ (Fig. 2(a)). The h-LuFeO$_3$ phase is expected to have higher X-ray absorption at 709 eV with *s*-polarized X-ray, corresponding to a brighter color (the "background") in Fig. 2(b). The dark "island" is the o-LuFeO$_3$ phase since the absorption is a local minimum. At 710 eV, the dark "islands" of the o-LuFeO$_3$ phase turn to bright as shown in Fig. 2(c) since the $t_{2g}$ peak of the o-LuFeO$_3$ phase dominates at 710 eV. To better distinguish the structural phases of LuFeO$_3$, we plot the absorption spectra (Fig. 2(e)), generated with *s*-polarized X-rays, along a sequence of positions, as shown in the X-PEEM image of Fig. 2(d). The spectra measured at position (7) is consistent with that of the o-LuFeO$_3$ phase, while the spectra measured at position (1) is consistent with that from the h-LuFeO$_3$ phase. A rapid change in the X-ray absorption spectra is observed between position (4) and (5), indicating a sharp structural interface.

The evidence of structural phase separation is also observed in the spectra that corresponds to excitation from the O K edge. Fig. 3 shows five X-ray absorption spectra at O K edge for both *s* and *p* polarization from sample region indicated in the dashed box in Fig. 2(c) (along the arrow of Fig. 2(c)). The spectra measured in the bright region in Fig. 2(c) indicate an o-LuFeO$_3$ electronic structure and while the spectra measured in the gray region indicate the h-LuFeO$_3$ electronic structure as discussed extensively.[18] The transition between the two structural domains is illustrated by spectra (2)-(4) in Fig. 3. This is the peak evolution observed at oxygen absorption edge.

In order to visualize more details of the h-LuFeO$_3$/o-LuFeO$_3$ interface, we scanned the surface morphology using atomic force microscopy on the thin sample used for the XAS studies, as shown in Fig. 4. There appear to be at least two different regions in the film: one flat and higher (with the surface closer to the tip), the other part more poorly defined, much rougher. The flatter regions are from the original h-LuFeO$_3$ phase film, while the rougher regions result from part of the film transformed into the o-LuFeO$_3$ phase. Fig. 4(b) shows well-defined steps (boundaries) separating the h-LuFeO$_3$ and o-LuFeO$_3$ phases (10 nm high). The angles between these boundaries are about 120º (illustrated by the dashed line in Fig. 4(b)). The boundary between the two structural phases appears to have a tendency to align with the crystal planes of the h-LuFeO$_3$ phase.

Our results indicate that the critical thickness for a stable h-LuFeO$_3$ phase on Al$_2$O$_3$ is actually smaller than 10 nm. The large interfacial energy at the boundary between the two LuFeO$_3$ phases appears the key to forming the large structural domains and the phase separation in LuFeO$_3$. The streaks visible in the images, in the region of o-LuFeO$_3$, are also about 10 nm in height; they occur at relative angles of 60º and are indicative of spatial movement of the structural domain wall, likely leaving defects in large number in specific locations to promote strain relief.

We have shown that the h-LuFeO$_3$ (001)/Al$_2$O$_3$ (0001) film is metastable even for a film thickness of 10 nm. On the other hand, the irreversible, 1$^{st}$ order transition from the h-LuFeO$_3$ phase to the o-LuFeO$_3$ phase requires an annealing temperature as high as 1000 °C, due to the large energy barrier to form the h-LuFeO$_3$/o-LuFeO$_3$ interface, suggesting no practical instability problems to retaining h-LuFeO$_3$, once grown, under normal (ambient) conditions. An important implication is that the previously measured ferroelectric to paraelectric transition at about 1050 K (~780 °C)[1] is not supposed to be affected by the instability significantly. Nevertheless, future investigations on the properties of h-LuFeO$_3$ films at elevated temperature need to be watchful of the emerging o-LuFeO$_3$ phase. The observation of the sharp, well-aligned boundaries between the hexagonal and orthorhombic phases, in a micrometer length scale, suggests the possibility of fabricating junctions between the two phases by self-organization, to better exploit this multiferroic h-LuFeO$_3$/o-LuFeO$_3$ (ferroelectric and antiferromagnetic) interface for nonvolatile magnetoelectric devices for spintronic applications.[21]

**Acknowledgements:**

This project was primarily supported by the National Science Foundation through the Nebraska Materials Research Science and Engineering Center (grant No. DMR-1420645). Additional support was provided by the Semiconductor Research Corporation through the Nanoelectronics Research Corporation (NERC), a wholly-owned subsidiary of the Semiconductor Research Corporation (SRC), through the Center for Nanoferroic Devices (CNFD), an SRC-NRI Nanoelectronics Research Initiative Center under Task ID 2398.001 and the National Science Foundation through grant ECCS – 1508541. The Canadian Light Source is funded by the Canada Foundation for Innovation, the Natural Sciences and Engineering Research Council of Canada, the National Research Council Canada, the Canadian Institutes of Health Research, the Government of Saskatchewan, Western Economic Diversification Canada, and the University of Saskatchewan.

**References**

[1] W. Wang, J. Zhao, W. Wang, Z. Gai, N. Balke, M. Chi, H.N. Lee, W. Tian, L. Zhu, X. Cheng, D.J. Keavney, J. Yi, T.Z. Ward, P.C. Snijders, H.M. Christen, W. Wu, J. Shen, and X. Xu, Phys. Rev. Lett. **110**, 237601 (2013).

[2] S.M. Disseler, J.A. Borchers, C.M. Brooks, J.A. Mundy, J.A. Moyer, D.A. Hillsberry, E.L. Thies, D.A. Tenne, J. Heron, M.E. Holtz, J.D. Clarkson, G.M. Stiehl, P. Schiffer, D.A. Muller, D.G. Schlom, and W.D. Ratcliff, Phys. Rev. Lett. **114**, 217602 (2015).

[3] J.A. Moyer, R. Misra, J.A. Mundy, C.M. Brooks, J.T. Heron, D.A. Muller, D.G. Schlom, and P. Schiffer, APL Mater. **2**, 012106 (2014).

[4] X. Xu and W. Wang, Mod. Phys. Lett. B **28**, 1430008 (2014).

[5] A.R. Akbashev, A.S. Semisalova, N.S. Perov, and A.R. Kaul, Appl. Phys. Lett. **99**, 122502 (2011).

[6] H. Das, A.L. Wysocki, Y. Geng, W. Wu, and C.J. Fennie, Nat Commun **5**, 2998 (2014).

[7] Y. Tokura and H.Y. Hwang, Nat Mater **7**, 694 (2008).

[8] N.A. Spaldin, S.W. Cheong, and R. Ramesh, Phys. Today **63**, 38 (2010).

[9] H. Schmid, Ferroelectrics **162**, 317 (1994).

[10] D. Treves, J. Appl. Phys. **36**, 1033 (1965).

[11] A.A. Bossak, I.E. Graboy, O.Y. Gorbenko, A.R. Kaul, M.S. Kartavtseva, V.L. Svetchnikov, and H.W. Zandbergen, Chem. Mater. **16**, 1751 (2004).

[12] E. Magome, C. Moriyoshi, Y. Kuroiwa, A. Masuno, and H. Inoue, Jpn. J. Appl. Phys. **49**, 09ME06 (2010).

[13] Y.K. Jeong, J. Lee, S. Ahn, and H.M. Jang, Chem. Mater. **24**, 2426 (2012).

[14] H. Wang, I. V. Solovyev, W. Wang, X. Wang, P.P.J. Ryan, D.J. Keavney, J.-W. Kim, T.Z. Ward, L. Zhu, J. Shen, X.M. Cheng, L. He, X. Xu, X. Wu, J. Shen, L. He, X. Xu, and X. Wu, Phys. Rev. B **90**, 014436 (2014).

[15] R.L. White, J. Appl. Phys. **40**, 1061 (1969).

[16] W. Wang, H. Wang, X. Xu, L. Zhu, L. He, E. Wills, X. Cheng, D.J. Keavney, J. Shen, X. Wu, and X. Xu, Appl. Phys. Lett. **101**, 241907 (2012).

[17] S. Cao, X. Zhang, T.R. Paudel, K. Sinha, X. Wang, X. Jiang, W. Wang, S. Brutsche, J. Wang, P.J. Ryan, J.-W. Kim, X. Cheng, E.Y. Tsymbal, P.A. Dowben, and X. Xu, J. Phys. Condens. Matter **28**, 156001 (2016).

[18] S. Cao, T.R. Paudel, K. Sinha, X. Jiang, W. Wang, E.Y. Tsymbal, X. Xu, and P. a Dowben, J. Phys. Condens. Matter **27**, 175004 (2015).

[19] E. Fermi, *Thermodynamics* (Dover Publications, New York, 1956).

[20] J. Stohr and H.C. Siegmann, *Magnetism from Fundamentals to Nanoscale Dynamics* (Springer, Berlin, 2006).

[21] P. Dowben, C. Binek, and D. Nikonov, in *Nanoscale Silicon Devices* (CRC Press, 2015), pp. 255–278.

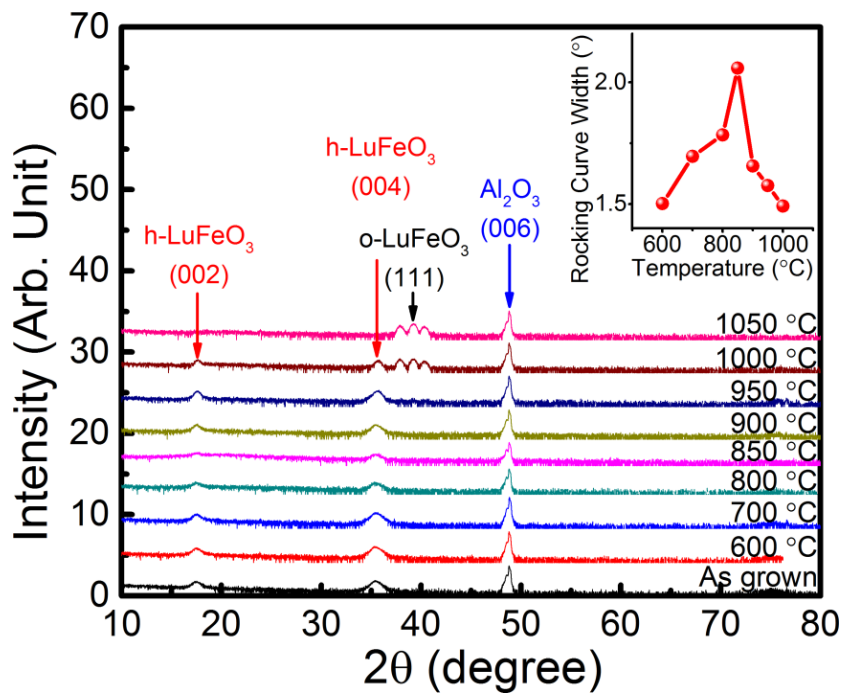

**Figure 1**. The θ-2θ X-ray diffraction spectra for a ~40 nm thick h-LuFeO$_3$ film grown on Al$_2$O$_3$, after being annealed at the stated temperatures. The inset is the rocking curve width of the h-LuFeO$_3$ (004) peak, as a functional of the annealing temperature $T_A$. The o-LuFeO$_3$ (111) peaks are labeled using the pseudo cubic indices. In orthorhombic structure, the (111) peak is, in fact, split into three peaks.

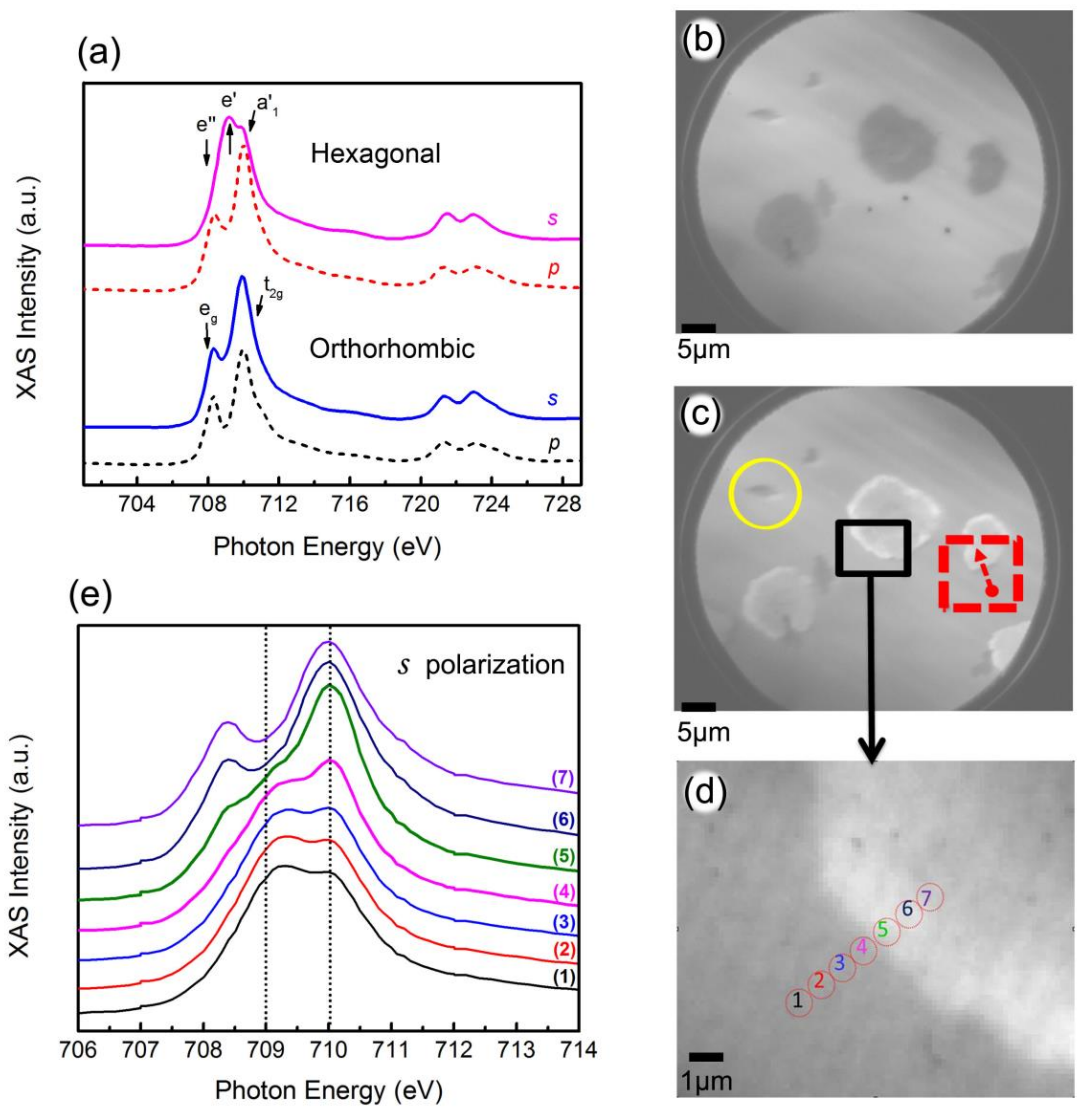

**Figure 2** The X-ray absorption spectra at Fe $L_{III}$ and $L_{II}$ edges and the associated X-PEEM images for a ~10 nm thick LuFeO$_3$ film grown on Al$_2$O$_3$. (a) The X-ray absorption spectra with *s* and *p* polarization for both h-LuFeO$_3$ and o-LuFeO$_3$. The PEEM image in a 50 µm field-of-view at 709 eV (b) and 710 eV (c) taken using *s*-polarized X-rays. The (yellow) circled region shows the morphological defects in the h-LuFeO$_3$ film. The (red) dash boxed region with arrow shows the starting point and direction of five oxygen K edge spectra illustrated in Fig. 3. The (black) boxed region in (c) was magnified into (d). (e) XAS

obtained corresponding to the seven circled positions in (d); the dash lines in (e) indicates the energy position of X-PEEM images taken at 709 eV (b) and 710 eV (c).

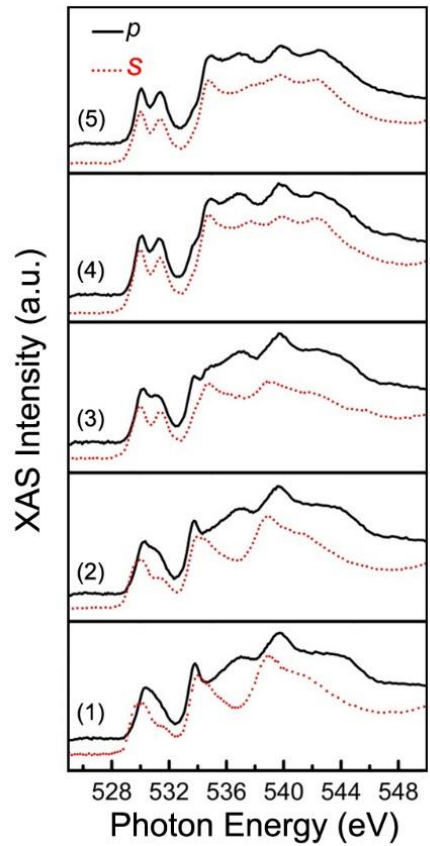

**Figure 3**. Five X-ray absorption spectra at O K edge starting from label (1) to (5) picked in the region indicated in Fig. 2(c) as the (red) dashed box.

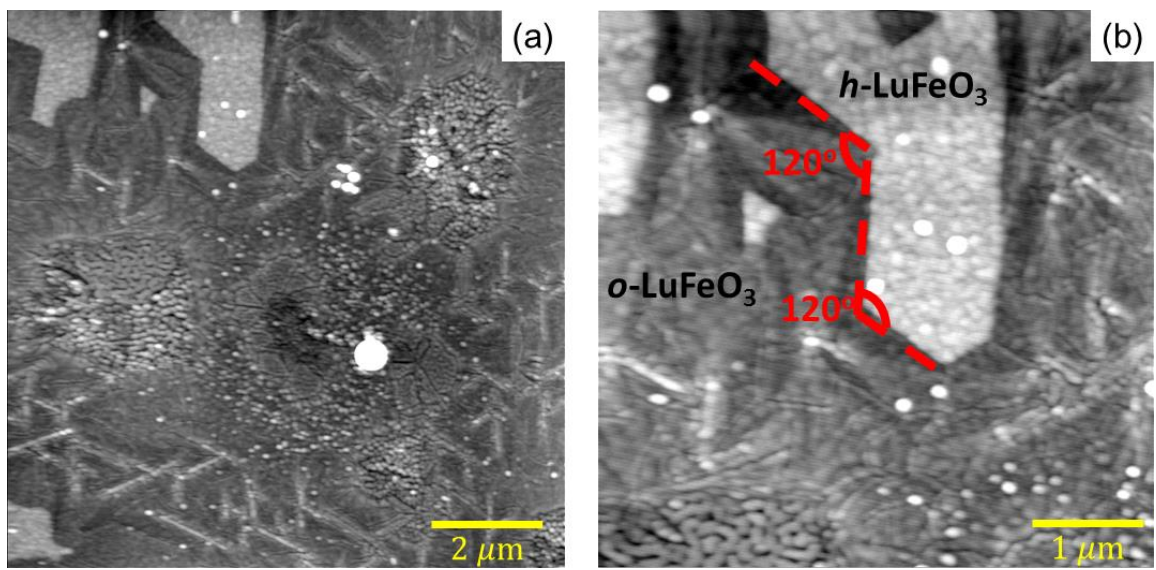

**Figure 4**. The atomic force microscopy (AFM) images illustrating the phase separation in a ~10 nm thick LuFeO$_3$ film grown on Al$_2$O$_3$. (a) The AFM image of a 10 μm×10 μm sample area. (b) The image of a 5 μm×5 μm sample area. The dashed (red) line in (b) shows the angle of 120° at boundary between hexagonal and orthorhombic phases.